# SPRING GRAVIMETER CALIBRATION EXPERIMENT WITH AN INERTIAL ACCELERATION PLATFORM

Ping Zhu[1], Michel van Ruymbeke[1], Francis Renders[1], Jean-Phillippe Noel

[1]Royal Observatory of Belgium, Avenue Circulaire 3, 1180, Brussels, Belgium, zhuping@oma.be, Fax: +32 2 3739822, Tel: +32 2 3730210

**Abstract**

Keywords: inertial acceleration platform, scale factor, gravimeter

*This experiment tested an automatically calibrate the relatively gravimeter with an absolute inertial force. The whole calibration system was controlled by microprocessor and low frequency oscillations were generated by a step motor. It could produce different period low frequency sinusoid inertial force. A LVDT (Linear Variable Differential Transformer) sensor was introduced to determine the frequency transfer function and the stability of platform vertical motion. The error of vertical displacement induced by the mechanical part of platform is less than 0.01 mm. Two LCR gravimeters (G336, G906) were settled on the platform. G336 was in the centre of platform and G906 was in the left side. A calibration program has been integrated in the micro processor which could send calibrate signal once a week. During the more than 7 moths experiment, the scale factor of G336 is 8128.647 nm/s²/v with ±1.1% uncertainties and G906 is 9421.017 nm/s²/v with ±2.7% fluctuations.*

**Introduction**

Absolute calibrate the relative gravimeter scale factor in the 0.1% precision is still an important issue. There a several calibration experiments can reach such a precision by side by side comparison with an absolute gravimeter[Francis, 1997; Iwano, et al., 2003], until now, the best results of the spring gravimeter calibration was get by comparison with a well know scale factor SGs [Francis and Hendrickx, 2001; Meurers, 2002]. No calibration can be operated remotely and automatically. This is the first time to automatically calibrate the relatively gravimeter with an absolute initial force.

The method to compute the scale factor of the two gravimeters is that the scale factor of gravimeter is directly computed from the slope of linear regression between the maximum acceleration values of platform with the readings of gravimeter. It will produce some bias if less calibration period, but inversely, it could induce gravimeter drift by a large and long calibration series. In order to obtain a stable scale factor in short term calibration, a new computation method is introduced in this paper. We use a zero phase Finite Impulse Response (zero phase FIR) digital filter to separate the calibration series from the earth tide and instruments drift before determining the scale factor. Then the scale factor is separately calculated by the linear regression between the whole set of gravimeter calibration records and the stacking results of same periodical records with synthetic platform accelerations.

**The initial acceleration platform**

The oscillating platform is consisted in a horizontal platform, long flat spring, three cylinders, eccentric wheel, cog-wheel, LVDT and step motor. A horizontal platform is resting on four vertical feet, each fixed at its lower extremity to a long flat spring which is rolling on three cylinders. A 1000-step stepping motor drive a large 100-teeth cog-wheel this will produce the rotation of eccentric plate. Then the vertical oscillation was generated [M. van. Ruymbeke et al., 1995]. The frequency was introduced by a 10 MHz quartz oscillator signal and sent to a programmable frequency divider controlled by a microprocessor. LVDT was installed in the platform to monitor the vertical displacement after it is calibrated by a LASER interferometer at Brussels before it was set up. Displacement of the platform could be expressed as

$$Z(t) = A_0 \cos(\omega t) \quad (1)$$

So that the corresponding acceleration produced by it is:

$$a = \frac{d^2 A}{dt^2} = -A_0 \omega^2 \cos(\omega t) \quad (2)$$

And the maximum acceleration is:

$$a_{max} = \pm A_0 \left(\frac{2\pi}{T}\right)^2 \quad (3)$$

Here, Z is vertical moving trace of platform; A0 is the maximum amplitude of vertical displacement of the platform A0=5.05mm; w is angular speed; t is time in second, T is calibration period. Three sets of different periodical motions were tested. At each set, we sent four calibration signals (Table1). The synthetic acceleration was plotted at Figure 1.

| I | | II | | III | |
|---|---|---|---|---|---|
| $T_0(s)$ | $a_{max}(nm/s^2)$ | $T_0(s)$ | $a_{max}(nm/s^2)$ | $T_0(s)$ | $a_{max}(nm/s^2)$ |
| 200 | ±4984.150 | 275 | ±2636.245 | 250 | ±3189.856 |
| 225 | ±3938.094 | 325 | ±1887.489 | 275 | ±2636.245 |
| 275 | ±2636.245 | 375 | ±1417.714 | 300 | ±2215.178 |
| 300 | ±2215.178 | 425 | ±1103.756 | 325 | ±1887.489 |

**Table1:** To find out the optimized period of the oscillation, three sets of period were tested in chronological order. I, the firstly tested, II the second and III the third.

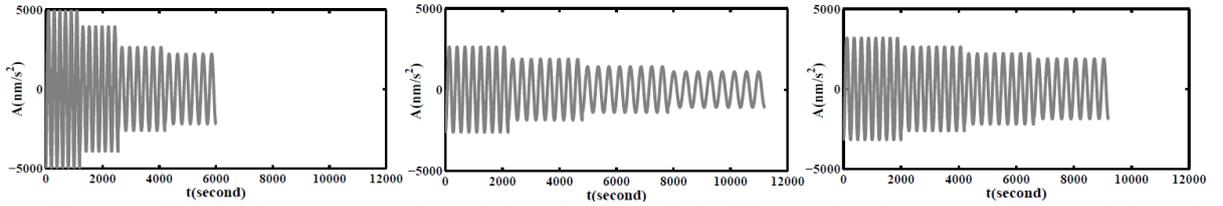

Figure 1 The synthetic acceleration of the platform. (Left) The accelerations were computed form the periods of the stage I; (Middle) The acceleration of II (Right) The third set of acceleration III.

The platform gravimeter calibration is usually to compute the scale factor directly by linear regression between maximum values of different acceleration values with the corresponding gravimeter record. The maximum acceleration was calculated in advance with equation (3). The relative gravimeter outputs volts or frequency could be obtained by the average readings during several cycles of one period calibration. For one period sinusoid acceleration, in principle the more cycles is run the more accurate gravimeter record could be reach. But, the errors from the nonlinearity influence term in equation (4) will also be increased if the calibration length was prolonged.

$$g_{cal} = g_{obs.} - g_T - g_{grad.} - g_\varepsilon \quad (4)$$

However, if the calibration period was too short, second derivative of vertical displacement produced by the short stroke of platform could induce a large effect on the gravimeter. In order to avoid these two kinds of errors; we should choice the suitable period of oscillation. The period selection was done by a systematically lift adjustment experiment at Brussels. The new mathematic computation model was also introduced. From equation (4) we can rebuild it as:

$$g_{obs.}(f_k) = \sum_{k=1}^{n}(\int g(t)e^{-j2\pi ft}dt) + g_{grad.} + g_\varepsilon \quad (5)$$

Here k is the kth harmonic wave of earth tide, f is the frequency of kth wave, t is time. The majority energy of earth tide is mainly rest on very low frequency band ($10^{-5}$ Hz for semidiurnal wave to $10^{-6}$ Hz for monthly wave). Then the calculation problem became a signal separation. The purely response of gravimeter to the acceleration force could be obtained, if we can separate it from the earth tides, instruments drift, and other nonlinearity factor such as climate effect.

The calibration and nonlinearity signal's frequency band is more higher compare to earth tide and the characterizes provide a possibility to separate low frequency earth tide signal and instrument drift from observation with a digital filter. In modern digital seismic acquisition systems, the zero phase Finite Impulse Response filters have been well developed and it's performance is systematically analyzed by Scherbaum [Scherbaum, 1997; Scherbaum and Bouin, 1997]. Zero phase digital filters are passing signals without phase changes, causing only a constant time shift. If this time shift is zero or is corrected for, the filter is called a zero phase filter. It could achieve maximum resolution of signal and no distortion of the input signal[Oppenheim and Schafer, 1999]. We applied the same technique to the gravimeter record. Two gravimeter's records were firstly filtered by a zero phase FIR filter to separate the low frequency signals, and then we subtract the observed data from it, finally the high frequency signals: calibration signal, background noise and seismic events were kept. The gravimeter record obtained from the zero phase FIR filter and synthetic acceleration was adjusted in same phase by cross correlation function[Carter, 1987; Quazi, 1981]. In Figure2, the gravimeter record and synthetic acceleration was plotted together.  Figure2 (left) shows the acceleration recorded by LCRG336.  The residuals were also plotted, the errors was reaching maximum when the acceleration of the platform was at its top value. We can stack the four periods in one to obtain a better resolution signal (Figure 2 right).

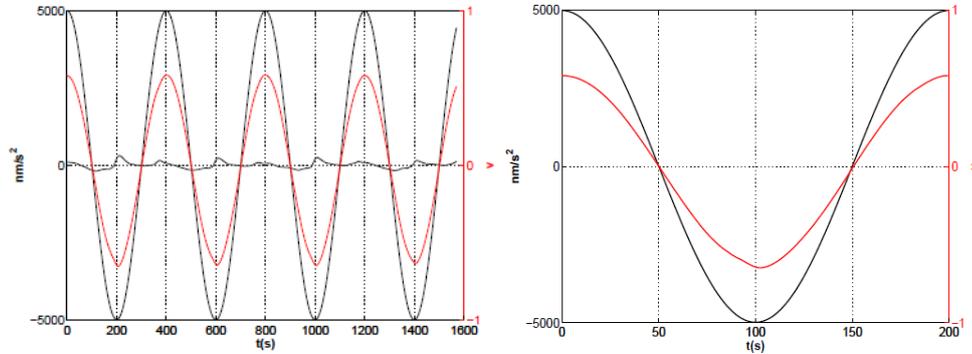

Figure 2, an example of the calibration signal recorded by the G336, (left) black line is the synthetic value and red one is the recording of G336. T0 = 200s; (right) is corresponding to the stacking results of each period.

The slope of linear model between the both gives directly absolute calibration of gravimeter. During the calibration, the periods were changed three times. For each period we only keep the oscillations at the middle stage, the first and last one were dropped. Two values were computed, one is from the separated calibration signs and the other is from the stacking results. We took the mean value of them as the final scale factor of one experiment (Table2).

**Conclusion and discussions**

Two gravimeters were calibrated by a sinusoid internal acceleration in situ and automatically at the Walfendange underground laboratory. The scale factor of G336 is 8128.647 nm/s$^2$/v with ±1.1% uncertainties and G906 is 9421.017 nm/s$^2$/v with ±2.7% fluctuations (Table 2) computed from 7 months experiment. Two scale factors with the maximum error of G906 were at the beginning of the experiment.  It was mainly due to the human interruption of set up. If only taking into account the most stable calibration from No.7 to No. 12, the variances of scale factor of G906 can reach the level of around ±0.2%. At No.20 the power supply was cut after that we readjusted the instruments.  The sensitive of the G906 changed a little but with a relative stable scale factor.  Since the mean value of scale factor is computed with complete set of the calibration, the last five calibrations have larger errors (Figure3).  The scale factor of G336 is more stable than the G906 but with an obvious nonlinear drift than G906.  If only select the most stable calibrations, we also can get the scale factor with relative small errors around ±0.2% from No.18 to No.22.

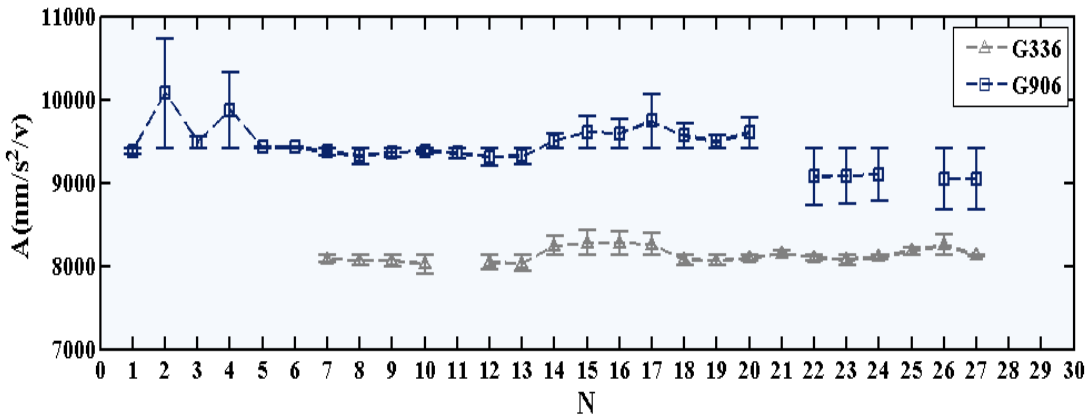

**Figure 3** The scale factor of the G336 is marked by the gray triangle and the scale factor of the G906 is marked by the blue rectangular. The error bar is the variance between the mean value and each individual measurement.

|     |        | G336    |        | G906     |        |
| --- | ------ | ------- | ------ | -------- | ------ |
| No. | Date   | s       | $\delta s$ | s        | $\delta s$ |
| 01  | Feb.♭  | -       | -      | 9380.82  | −0.43% |
| 02  | Feb.♭  | -       | -      | 10070.57 | 6.90%  |
| 03  | Mar.♭  | -       | -      | 9483.61  | 0.66%  |
| 04  | Mar.♭  | -       | -      | 9873.88  | 4.81%  |
| 05  | Apr.♭  | -       | -      | 9421.65  | 0.01%  |
| 06  | Apr.♭  | -       | -      | 9428.30  | 0.08%  |
| 07  | May.♭  | 8082.27 | −0.57% | 9373.77  | −0.50% |
| 08  | May.♭  | 8068.32 | −0.74% | 9316.05  | −1.11% |
| 09  | May.♭  | 8062.94 | −0.81% | 9362.29  | −0.62% |
| 10  | May.♭  | 8019.26 | −1.35% | 9382.08  | −0.41% |
| 11  | May.♭  | -       | -      | 9354.97  | −0.70% |
| 12  | Jun.♭  | 8038.47 | −1.11% | 9307.41  | −1.21% |
| 13  | Jun.♭  | 8033.58 | −1.17% | 9321.17  | −1.06% |
| 14  | Jun.♭  | 8241.01 | 1.38%  | 9501.34  | 0.85%  |
| 15  | Jun.♭  | 8278.46 | 1.84%  | 9605.81  | 1.96%  |
| 16  | Jun.♭  | 8273.54 | 1.78%  | 9587.84  | 1.77%  |
| 17  | Jul.♭  | 8258.59 | 1.60%  | 9739.06  | 3.38%  |
| 18  | Aug.♮  | 8072.85 | −0.69% | 9565.89  | 1.54%  |
| 19  | Aug.♮  | 8068.37 | −0.74% | 9496.15  | 0.80%  |
| 20  | Aug.♮  | 8090.76 | −0.47% | 9599.91  | 1.90%  |
| 21  | Aug.♮  | 8152.19 | 0.29%  | -        | -      |
| 22  | Sep.♮  | 8098.66 | −0.37% | 9073.10  | −3.69% |
| 23  | Sep.♮  | 8071.72 | −0.70% | 9086.80  | −3.55% |
| 24  | Sep.♮  | 8107.30 | −0.26% | 9099.93  | −3.41% |
| 25  | Sep.♮  | 8176.46 | 0.59%  | -        | -      |
| 26  | Sep.♮  | 8253.70 | 1.54%  | 9045.23  | −4.00% |
| 27  | Sep.♮  | 8124.49 | −0.05% | 9047.80  | −3.96% |

**Table2:** ♭ the periods are from I, ♮ from III.

To accurately determine the scale factor of the relative gravity meter is still a challenge. The experiment made at the Walferdange laboratory is managed to study the behaviour of the relative gravimeter of G336 and G906 and try to find one simple alternative solution for the problem. Despite the characterization of the instrument itself, to increase the accuracy of the scale factor determination with an acceleration platform, several effects should be carefully studied.

a) The maximum vertical displacement of the acceleration platform should be checked before and after the experiment;

b) Any tilt of the platform during its vertical motion should be minimized;

c) The period of the oscillation must be controlled as accurate as possible; several seconds shift of the period will induce a significant amount of the unpredicted acceleration.

**Acknowledgements**

The experiment won't be possible without the support and hospitality of Dr. Olivier Francis. Thanks to Mr. Gilles Celli of ECGS who has prepared the original data file. The project is supported by Action 2 funds of Belgian Ministry of Scientific Politics.